\documentclass[fleqn,usenatbib]{mnras}

\usepackage{newtxtext,newtxmath}
\usepackage[T1]{fontenc}
\usepackage{ae,aecompl}

\usepackage{graphicx}
\usepackage{amsmath}
\usepackage{amssymb}

\usepackage[english]{babel}
\usepackage[separate-uncertainty,binary-units=true]{siunitx}
\usepackage{physics}
\usepackage[usenames,dvipsnames,svgnames]{xcolor}
\usepackage{url}
\usepackage{cleveref}
\usepackage{pdfpages}
\usepackage{placeins}

\usepackage{extdash}

\newcommand*\mean[1]{\bar{#1}}

\DeclareMathOperator{\ivar}{ivar}

\newcommand{\unsim}{{\mathord{\sim}\mkern1mu}}

\newcommand{\ComovingParallel}{D_C}

\newcommand{\rpar}{{r_\parallel}}
\newcommand{\rtrans}{{r_\perp}}

\newcommand{\LymanA}{Lyman\Hyphdash*$\upalpha$}
\newcommand{\LymanABold}{Lyman\Hyphdash*$\boldsymbol{\upalpha}$}

\newcommand{\LymanB}{Lyman\Hyphdash*$\upbeta$}

\newcommand{\CaHK}{{\ion{Ca}{ii}~H\&K}}

\newcommand{\SDSSFluxUnits}{\erg\per\second\per\centi\meter^2\per\angstrom}

\DeclareSIUnit\year{yr}
\DeclareSIUnit\hubbleparameter{\textit{h}}
\DeclareSIUnit\hprecip{\hubbleparameter^{-1}}
\DeclareSIUnit\parsec{pc}
\DeclareSIUnit\mpc{\mega\parsec}
\DeclareSIUnit\hrmpc{\hprecip\mega\parsec}
\DeclareSIUnit\erg{erg}

\newlength\figureheight
\newlength\figurewidth

\crefname{appsec}{Appendix}{Appendices}

\newcommand{\AckSDSSIII}{Funding for SDSS-III has been provided by the Alfred P. Sloan Foundation, the Participating Institutions, the National Science Foundation, and the U.S. Department of Energy Office of Science. The SDSS-III web site is \url{www.sdss3.org}.\par SDSS-III is managed by the Astrophysical Research Consortium for the Participating Institutions of the SDSS-III Collaboration including the University of Arizona, the Brazilian Participation Group, Brookhaven National Laboratory, Carnegie Mellon University, University of Florida, the French Participation Group, the German Participation Group, Harvard University, the Instituto de Astrof\'isica de Canarias, the Michigan State/Notre Dame/JINA Participation Group, Johns Hopkins University, Lawrence Berkeley National Laboratory, Max Planck Institute for Astrophysics, Max Planck Institute for Extraterrestrial Physics, New Mexico State University, New York University, Ohio State University, Pennsylvania State University, University of Portsmouth, Princeton University, the Spanish Participation Group, University of Tokyo, University of Utah, Vanderbilt University, University of Virginia, University of Washington, and Yale University.}

\title[ISM absorption in the Lyman\Hyphdash*$\alpha$ forest]{The Effect of Interstellar Absorption on Measurements of the Baryon Acoustic Peak in the Lyman\Hyphdash*$\boldsymbol{\upalpha}$ Forest}

\author[Vadai et al.]
{Yishay Vadai$^{1}$,
Dovi Poznanski$^{1}$,
Dalya Baron$^{1}$,
Peter E. Nugent$^{2}$,
\newauthor and David Schlegel$^{3}$
\\
$^{1}$School of Physics and Astronomy, Tel-Aviv University, Tel Aviv 69978, Israel\\
$^{2}$Computational Cosmology Center, Lawrence Berkeley National Laboratory, 1 Cyclotron Road, Berkeley, CA 94720, USA\\
$^{3}$Physics Division, Lawrence Berkeley National Laboratory, 1 Cyclotron Road, Berkeley, CA 94720, USA
}

\date{Accepted XXX. Received YYY; in original form ZZZ}

\pubyear{2017}

\begin{document}
\label{firstpage}
\pagerange{\pageref{firstpage}--\pageref{lastpage}}
\maketitle

\begin{abstract}
In recent years, the autocorrelation of the hydrogen \LymanA{} forest has been used to observe the baryon acoustic peak at redshift $2<z<3.5$ using tens of thousands of QSO spectra from the BOSS survey. However, the interstellar medium of the Milky-Way introduces absorption lines into the spectrum of any extragalactic source. These lines, while weak and undetectable in a single BOSS spectrum, could potentially bias the cosmological signal. In order to examine this, we generate absorption line maps by stacking over a million spectra of galaxies and QSOs. We find that the systematics introduced are too small to affect the current accuracy of the baryon acoustic peak, but might be relevant to future surveys such as the Dark Energy Spectroscopic Instrument (DESI). We outline a method to account for this with future datasets. 
\end{abstract}




\section{Introduction}

The baryon acoustic peak is a cosmological feature of a predictable physical size, which can be used to constrain cosmological parameters that describe the expansion of the universe.
It was first discovered in the power spectrum of temperature differences in the Cosmic Microwave Background (CMB) radiation, where it is known as Baryon Acoustic Oscillations (BAO; \citealt{2000Natur.404..955D},  \citealt{0004-637X-743-1-28,1475-7516-2013-10-060,2013ApJS..208...19H,2014A&A...571A..16P}).
The same phenomenon can be seen as a single peak in the autocorrelation of the baryon density, and it has first been observed in the autocorrelation of galaxy number counts \citep{2005ApJ...633..560E}.

The Sloan Digital Sky Survey (SDSS; \citealt{2000AJ....120.1579Y}) has so far obtained spectra for about \SI{2.5e6}{} galaxies, spread over about a third of the sky. As part of SDSS, the Baryon Oscillation Spectroscopic Survey (BOSS; \citealt{2013AJ....145...10D}), has observed almost \SI{3e5}{} QSOs \citep{paris17}. In recent years, there have been a number of works which used BOSS data to measure the baryon peak using absorption in the \LymanA{} forest of QSOs at a redshift range of about $2<z<3.5$ \citep{2013A&A...552A..96B,1475-7516-2013-04-026,2015A&A...574A..59D}.

The baryon acoustic peak is a weak feature, and the typical signal-to-noise ratio (SNR) of the \LymanA{} forest is low in an individual BOSS spectrum. The aggregate signal from over $\unsim{}10^4$ QSOs is required in order to achieve the required SNR in the autocorrelation function. This makes this process potentially sensitive to systematics that could be in the data, even if they cannot be detected in a single QSO spectrum. In this paper, we focus on one such effect, which is intervening absorption lines formed in the interstellar medium (ISM) of the Milky Way (MW). Interstellar emission (see e.g., \citealt{2012ApJ...744..129B}) will not be considered here. 

ISM absorption at some wavelengths, if not accounted for, could mimic \LymanA{} at particular redshifts. If there are correlations between absorption lines, or between sight lines, they could introduce a spurious signal to the 2D autocorrelation function of the \LymanA{} forest, and bias the shape or position of the baryon acoustic peak. This possible effect was first noted in \citet{lee13}, who provided a global correction for the spectra.

\citet{2012MNRAS.426.1465P} followed by \citet{2015MNRAS.447..545B,2015MNRAS.451..332B}, have used the integrated MW absorption spectrum in order to study correlations of absorption features with dust extinction, and the properties of the Diffuse Interstellar Bands (DIBs; see also \citealt{murga15} and \citealt{lan15}). They achieved this by stacking hundreds of thousands of SDSS spectra, in which the lines are not detected individually. The SDSS coverage allowed them to probe different extinction regimes, as well as different directions on the sky. In this work we use the same methodology in order to measure the ISM features. In order to calculate the autocorrelation of the \LymanA{} forest, we re-created a pipeline which is similar to previous works  \citep{2013A&A...552A..96B,1475-7516-2013-04-026,2015A&A...574A..59D}. Combining these two approaches, we study the ISM contribution in isolation from other effects.

Near the final stages of this work, \citet{2017arXiv170200176B} published an updated measurement of the \LymanA{} BAO autocorrelation from 12th data release (DR12) of SDSS BOSS \citep{2015ApJS..219...12A}. There, they briefly discuss the possible systematic contribution of the MW ISM to the BAO signal, which we study here in more detail.

\section{Baryon Peak Pipeline}
In order to measure the ISM contribution to the autocorrelation of the \LymanA{} forest, we need a pipeline to calculate the autocorrelation estimator.
We follow the method used in \citet{2013A&A...552A..96B}, \citet{1475-7516-2013-04-026}, and \citet{2015A&A...574A..59D}, with some modifications as discussed below.

\subsection{Data selection}
We use \SI{176944}{} QSOs from BOSS DR12 with redshift $2.1<z<3.5$ as selected by \citet{2012ApJS..199....3R}, and observed as part of the BOSS program with the upgraded SDSS spectrographs \citep{2013AJ....146...32S}. This redshift range translates to a wavelength range of about \SIrange{3600}{5500}{\angstrom}. The lower limit is set by atmospheric absorption of UV light, whereas the higher limit is set by diminishing observed QSO numbers at higher redshifts. Note that this wavelength range falls within the blue cameras of the BOSS spectrographs. In the rest frame, we restrict ourselves to the range \SIrange{1040}{1200}{\angstrom}. 

Within the spectra we discard pixels as follows. First, SDSS provides a variety of flags that are derived from the subexposures, and then propagated to the reduced stacked spectrum with both AND and OR operators. The OR operator corresponds to pixels that are marked in a single sub exposure, while the AND operator corresponds to pixels that are marked in all the sub exposures of a given source.We discard all pixels that are flagged in all sub-exposures, but we use only specific OR mask flags. Out of those, the \texttt{BRIGHTSKY} flag is the most common, and causes the removal of \SI{18}{} percent of the data. A few other flags (\texttt{BADTRACE}, \texttt{BADFLAT}, \texttt{MANYBADCOLUMNS}) discard up to 2 additional percent. We also mask the telluric \ion{Hg}{i} line with a \SIrange{4356}{4364}{\angstrom} mask. This line is flagged by the SDSS pipeline in most of the spectra, but we find some residuals in other spectra as well. 

Broad Absorption Lines (BALs) in the rest frame of the QSOs overlap with the \LymanA{} forest and introduce noise that we wish to avoid. We mask BALs in our spectra using the BOSS catalogue \citep[\texttt{DR12Q\_BAL}]{paris17}.
We mask out the following lines: \ion{C}{iv}, \ion{O}{vi}, \ion{NV}{}, \ion{Si}{iv}+\ion{O}{iv} as well as \LymanA{}, based on the supplied BAL reference frames (minimum and maximum velocities).
We add a fractional safety margin of \SI{2e-3}{} of the wavelength to the line widths. For the same reason, we mask Damped \LymanA{} Absorption systems (DLAs), using a catalogue provided by \citet{2016arXiv160504460G}. 
Similarly to \citet{2015A&A...574A..59D},
we remove regions with more than \SI{20}{} percent absorption and correct the damping wings using a Lorentzian profile for \LymanA{} and \LymanB{}, ignoring possible metal lines from these systems.

\subsection{Continuum fit}
We implement a continuum fit process based on MF-PCA (see \citealt{0004-637X-618-2-592,2011A&A...530A..50P,1538-3881-143-2-51}),  with some minor differences in the fit process.  First, we do not perform a visual inspection of all the fitted continua (except for a small fraction of objects).
\citet{1538-3881-143-2-51} fitted redshift correction, power-law correction and flux normalization parameters simultaneously with the PCA coefficients, using the Levenberg-Marquardt (LM) algorithm. We perform LM on power-law correction and flux normalization, with a linear least squares for the PCA coefficients in every iteration. We find that allowing the redshift to vary sometimes produces unreliable results.
For every QSO, we calculate a `goodness of fit' value, defined as the mean of the absolute fit errors, over a wavelength range of \SIrange{1216}{1600}{\angstrom}.
We bin the QSOs according to their SNR, and discard those with the lowest goodness of fit in every bin. To account for the lack of visual inspection, we discard the worst \SI{10}{} percent of the continuum fits, instead of the \SI{5}{} percent discarded in \citet{1538-3881-143-2-51}. Assuming the number of pairs is quadratic with QSO density, we thus remove about 20 percent of the pixel pairs.
At the end of this stage we discard 
29662 QSOs due to bad fit, 5967 QSOs in which the continuum level is comparable to the noise, and 546 QSOs in which the relevant part of the spectrum contains less than 150 valid pixels.
The \LymanA{} forest autocorrelation is computed from the remaining 140769 QSOs. This is a similar number to the total used by \citet{2015A&A...574A..59D} who used DR11 while we use DR12, but due to our more conservative cuts, and lack of visual inspection, we reject more QSOs.

\subsection{The autocorrelation estimator}
\label{sec:cor_est}

We obtain the transmittance, $T(z)$ for every QSO, where $z$ is the \LymanA{} equivalent redshift of the wavelength, $f(z)$ is the QSO intrinsic flux, and $C_q(z)$ is the estimated continuum.
\begin{equation}
T(z)=\frac{f(z)}{C_q(z)}
\end{equation}
$\bar{T}(z)$ is the mean transmittance, where the average is done over all QSOs, as a function of redshift.
The fractional change in transmittance, $\delta_F$, is defined as:
\begin{equation}
\delta_F(z) = \frac{T(z)}{\bar{T}(z)} - 1
\end{equation}

The overall slope predicted by the continuum fit is not exact, especially for higher redshift QSOs, where the \LymanA{} forest spans a wider wavelength range.
We define $\delta_F(\ComovingParallel(z))$ to be $\delta_F(z)$ as a function of the line-of-sight comoving distance, $\ComovingParallel(z)$.
We detrend $\delta_F(\ComovingParallel(z))$ of \LymanA{} forests that span more than \SI{500}{\mpc} ($h_0=0.7$), using a weighted nonuniform boxcar filter with a window size of $\SI{\pm300}{\mpc}$. We find, by trial and error, that the window size is large enough to avoid affecting the 2D autocorrelation in a way that distorts the baryon peak.

After calculating $\delta_F$, for all QSOs, following \citet{2013A&A...552A..96B}, we remove a residual bias, $\left<\delta_F(z)\right>$, in every redshift bin. 
We then calculate the autocorrelation estimator: 
\begin{equation}
\hat\xi_A = \frac{\sum_{ij\in A}w_{ij}\delta_i\delta_j}
{\sum_{ij\in A}w_{ij}}
\end{equation}
where A is a bin of comoving parallel and transverse distances between pairs of pixels. The sum over $i,j$ represents all pixel pairs with the required distances.

\subsection{Pixel weights}
\label{sec:pixel_weights}

We use a model for the weights given by \citet{2013A&A...552A..96B}, based on \citet{2006ApJS..163...80M}. The expression for the weight of a pair of pixels is:
\begin{equation}
w_{ij}\propto\frac{\left[(1+z_i)(1+z_j)\right]^{\gamma/2}}{\xi_{ii}^2\xi_{jj}^2}
\end{equation}

Where:
\begin{equation}
\xi_{ii}^2 = \frac{\sigma_{\text{pipeline},i}^2}{\eta(z_i)} + \sigma_\text{LSS}^2(z_i)\quad;\quad
\sigma_{\text{pipeline},i}^2 = \frac{1}{\ivar\cdot(C_q\mean{T})^2}
\end{equation}

This model takes into consideration:
\begin{itemize}
\item The measured redshift dependence of the correlation function $(1+z)^{\gamma}$ where $\gamma=3.8$.
\item The flux error for each point in the \LymanA{} forest $\sigma_{\text{pipeline},i}^2$, which in turn is based on flux error ($\ivar$).
\item Variance introduced by the large scale structure $\sigma_\text{LSS}$.
\item A scaling factor for the contribution of pipeline variance to the correlation variance $\eta$.
\end{itemize}

We use the values obtained by \citet{2013A&A...552A..96B} for $\sigma_{LSS}(z)$ and $\eta(z)$.

The contributions of $i$ and $j$ can be separated to the product $w_{ij} = w_i w_j$, so that weights can be calculated per pixel rather than for every pair:
\begin{equation}
w_i = (1+z_i)^{\gamma/2}\left[\dfrac{1}{\ivar\cdot\left(C_q \mean{T}_i\right)^2\eta(z_i)}+\sigma_\text{LSS}^2(z_i)\right]^{-1}
\end{equation}

\subsection{Implementation}
Our pipeline is written in Python+NumPy, with the critical path in the autocorrelation calculation written in Cython. We use the reference Python implementation, which is a bytecode interpreter, and as such it can be 2--3 orders of magnitude slower than compiled languages such as C. Each instruction has to be translated at runtime, which incurs an overhead.
Numpy offers improved performance by including pre-compiled versions of common mathematical calculations, like matrix operations.
However, some problems do not map efficiently to the `recipes' provided by Numpy.
Cython uses a simple syntax similar to python, while achieving performance that is comparable to C.
Using Cython, we can write loops explicitly and control the iteration boundaries, which allows us to ignore pixel pairs that are too far away in redshift.

\begin{figure} 
\begin{center}
\linespread{1.1}
\selectfont
\renewcommand{\AA}{\text{\r{A}}}
\setlength\figureheight{8cm}
\setlength\figurewidth{\minof{\linewidth}{18cm}}
\includegraphics[width=\figurewidth]{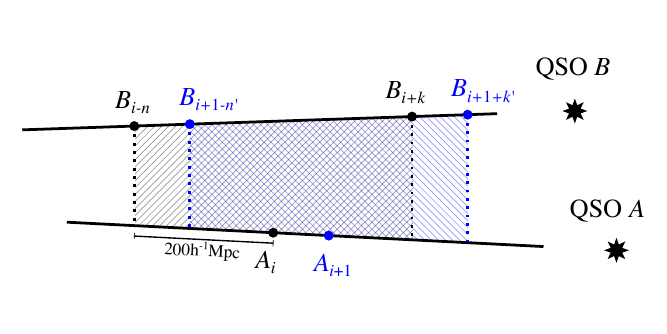}
\caption{A diagram of two sightlines $A$,$B$ showing how the range of relevant pixels in $B$ for every pixel in $A$, is determined by the maximum correlation length of interest.
When the autocorrelation length is small enough compared to a typical \LymanA{} forest length, it is possible to avoid iterating over all pairs of  pixels between two QSOs. The minimum (maximum) indices for each valid range form a monotonically increasing sequence. If we have a valid starting point $B_{i-n}$ for $A_i$, we can begin the next iteration $A_{i+1}$ at that point. At the other end, once at least one point in $B$ has been found to be in range of $A_i$, iteration over points in $B$ can stop if the next point is out of range. \label{fig:calc_corr}}
\end{center}
\end{figure} 

Given two QSOs, $A$ and $B$, with \LymanA{} forest sizes $N$ and $M$, respectively, calculating their correlation contribution using matrix operations requires all $N\times M$ combinations of pixel pairs to be computed.
A typical sightline can have a range of \SIrange{200}{600}{\hrmpc}, while the total range $1.95<z<3.5$ corresponds to \SI{1700}{\hrmpc}. Given a point in QSO A, the matching points within \SI{200}{\hrmpc} in QSO B all lie in a limited distance range. Therefore, the overlap between sightlines can be small, and most of the matrix elements represent points that are too far apart.
Because distances are monotonically increasing, it is possible to choose the iteration boundaries of the inner loop (over QSO B) to avoid pairs which cannot be found within the required range of \SI{200}{\hrmpc} (see \cref{fig:calc_corr}).
Suppose we know the maximal range of pixels in QSO B, $\left[B_{i-n}, B_{i+k}\right]$ that are in range of $A_i$.
It is possible to find the maximal range $\left[B_{i+1-n'}, B_{i+1+k'}\right]$ corresponding to the next point in $A$, $A_{i+1}$, without iterating over all points in $B$ for every $A_i$, because the series of range start (or end) points is monotonically increasing as well.

In order to find the set of all QSO pairs, we choose QSOs that are separated by less than \ang{3;16;} which corresponds to about \SI{200}{\hrmpc} at $z=1.9$.
The \SI{1.5e5}{} QSOs form about \SI{2e10}{} potential pairs, out of which only \SI{5e7}{} unique pairs are within the required angular range.
We use a KD-Tree implementation in the \emph{astropy} package, which can perform a k-Nearest Neighbor Search in $O(k \log n)$.
To parallelize the search, each core is assigned a different subset of the QSOs, to check against the full list of QSOs.

The pipeline supports parallelization through MPI, using the Mpi4Py package.
Performance is about \SI{3e7}{} pixel pairs per second per core. The BOSS \LymanA{} forest yields $\unsim\SI{3e12}{}$ pixel pairs and can be calculated within an hour on a 32-core machine.

Every pixel pair in our pipeline is binned according to sky region (HealPix bin), parallel comoving separation, transverse comoving separation, and parallel comoving distance (the average distance of the pixel pair). The range and resolution on each axis can be changed to obtain different cuts of the results (a tradeoff is required to prevent the size of the 4-dimensional array from becoming too large).

We make the pipeline source code available at:\\ \url{https://github.com/yishayv/lyacorr}. 

\subsection{\LymanABold{} autocorrelation results}
\begin{figure*} 
\renewcommand{\AA}{\text{\r{A}}}
\begin{center}
\includegraphics[width=\linewidth]{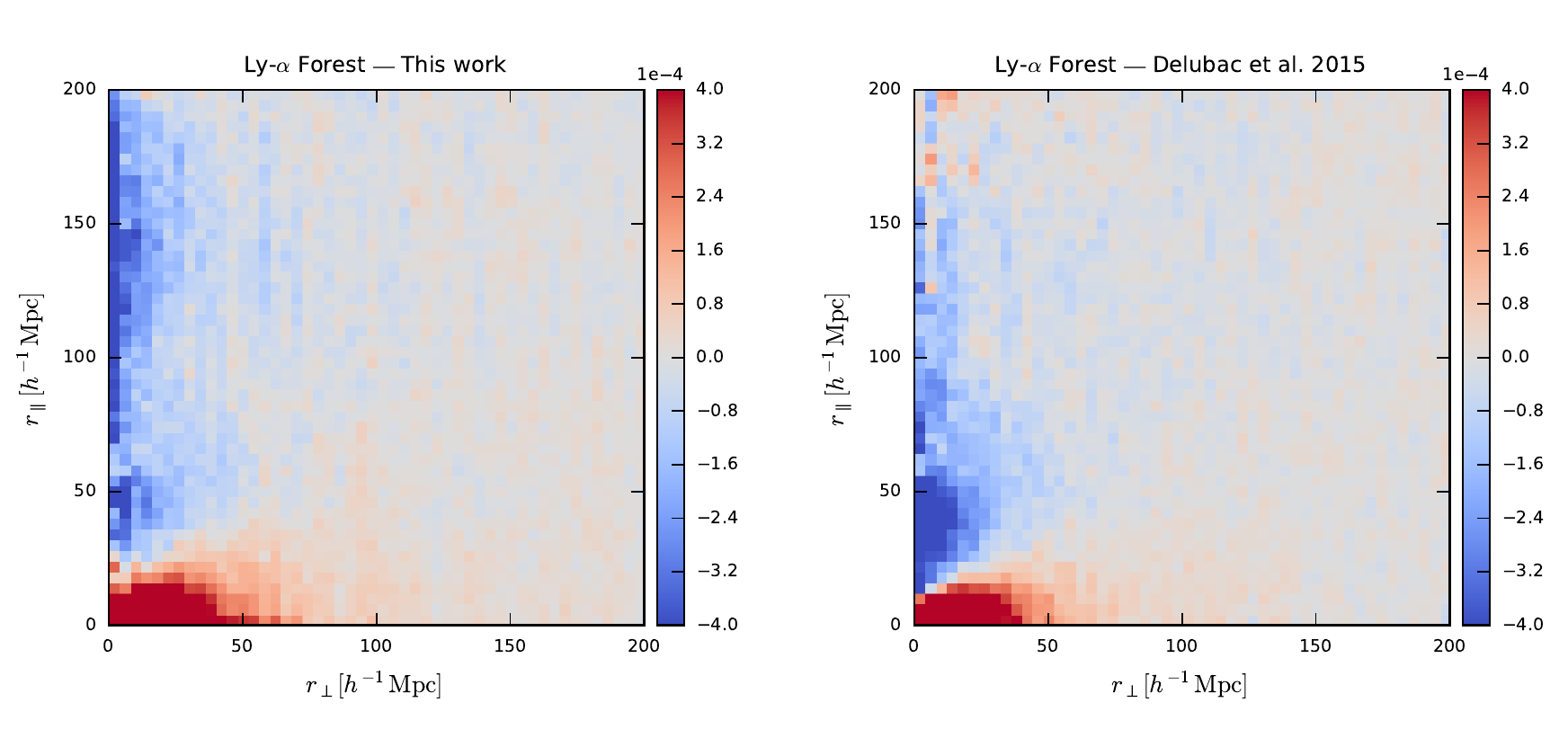}
\end{center}
\caption{Autocorrelation, $\hat{\xi}_A$, of the change in transmittance $\delta_F$, in comoving distance bins in parallel and transverse directions. The baryon peak should have the form of a quarter circle around the origin with a radius of $\unsim\SI{100}{\hrmpc}$.
In addition to a faint trace of the BAO signal, one can see a strong peak near the origin, due to the large scale structure, as well as spurious anticorrelation at short perpendicular distances due to redshift distortion \citep{1987MNRAS.227....1K,2011ApJ...728...34T}.
Left: This work.
Right: Data from \citet{2015A&A...574A..59D}.
\label{fig:corr_2d}}
\end{figure*}

\begin{figure} 
\renewcommand{\AA}{\text{\r{A}}}
\includegraphics[width=\columnwidth]{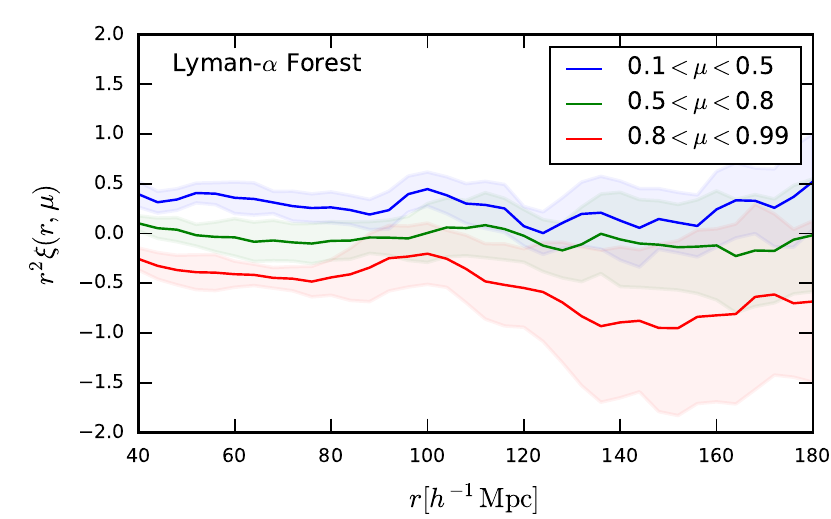}
\caption{\LymanA{} autocorrelation, $\hat{\xi}$, multiplied by distance squared in three angle ranges where $\mu$ is the cosine of the angle of the coordinate $(\rpar,\rtrans)$. We expect to see the acoustic peak at $\unsim\SI{100}{\hrmpc}$. Such a peak is only notable (if barely) at the smallest angles, similarly to the results of \citet{2015A&A...574A..59D}. 
Note the plotted autocorrelation and uncertainty do not take into account the covariance of the correlation bins. 
\label{fig:corr_mu}}
\end{figure}

\Cref{fig:corr_2d} shows the autocorrelation estimator, $\hat{\xi}$, as a function of parallel and transverse comoving separation, $(\rpar,\rtrans)$.
We include the autocorrelation estimator from \citet{2015A&A...574A..59D} for comparison (received by private communication).
\Cref{fig:corr_mu} shows the autocorrelation estimator multiplied by distance squared in three angular wedges where $\mu$ is the cosine of the angle of the coordinate $(\rpar, \rtrans)$.
There is faint trace of the BAO signal, at small angles (blue line), with an amplitude of about $\unsim\SI{5e-5}{}$ near $\SI{100}{\hrmpc}$.

Since our purpose is not to perform a full cosmological measurement, we did not derive the covariance matrix of the autocorrelation estimator, so we do not have a robust error estimation, in contrast to previous works. However, the autocorrelation estimator values are similar to previous results with the same data, with slightly more noise along $\rtrans$, appearing as faint vertical stripes in \cref{fig:corr_2d}. We can therefore proceed to study the ISM imprint on the cosmological signal. 

\section{Stacked ISM Spectra}
\citet{2012MNRAS.426.1465P} have used SDSS spectra stacked in the observer frame to measure the correlation of dust extinction with the \ion{Na}{i}\,D absorption doublet in the ISM of the MW. \citet{2015MNRAS.447..545B,2015MNRAS.451..332B} have expanded that process to the entire wavelength range of the SDSS spectrograph to detect multiple DIBs,  study their properties, and measure the correlations between DIBs and with other species. 

\citet{2015MNRAS.447..545B} used only data redder than \SI{3800}{\angstrom}, corresponding to a \LymanA{} redshift of 2.1, due to the limits of the original SDSS spectrographs \citep{2013AJ....146...32S}. We create ISM spectra using a similar process, using only spectra from the  BOSS spectrograph, with a limit of \SI{3600}{\angstrom}, or a \LymanA{} redshift below 2. We use  \SI{1.5e6}{} galaxies and \SI{3e5} QSOs from the 12th data release (DR12).
For every spectrum, we estimate the continuum using a Savitzky-Golay filter with a \SI{80}{\angstrom} window \citep{savitzky64}\footnote{Savitzky-Golay is a generalization of `running mean' where instead of fitting a constant in the window (the mean) one fits a polynomial.}. Filters have edge effects, they are ill defined near the edges of a sequence (spectrum in our case), over a region proportional to the window size. We use a narrower window than in the previous works, in order to discard less blue pixels, our region of interest. For every object (galaxy or QSO), we divide the original spectrum by the continuum to obtain a (noisy) detrended spectrum of relative absorption.
Pixels where the continuum was nonpositive or comparable to noise (below \SI{0.5e-17}{\SDSSFluxUnits}) were discarded. 

We then calculate the median of the detrended spectra, at every wavelength, in 20 bins of extinction, using the maps of \citet{1998ApJ...500..525S}. The bin boundaries are chosen so that there is a similar number of spectra in each bin. This stacking allows us to reach the SNR necessary to detect the weak absorption features we seek. As discussed below, these stacks allow us to address only some of aspects of the ISM  bias on the BAO measurement. 

\begin{figure}
\renewcommand{\AA}{\text{\r{A}}}
\includegraphics[width=\linewidth]{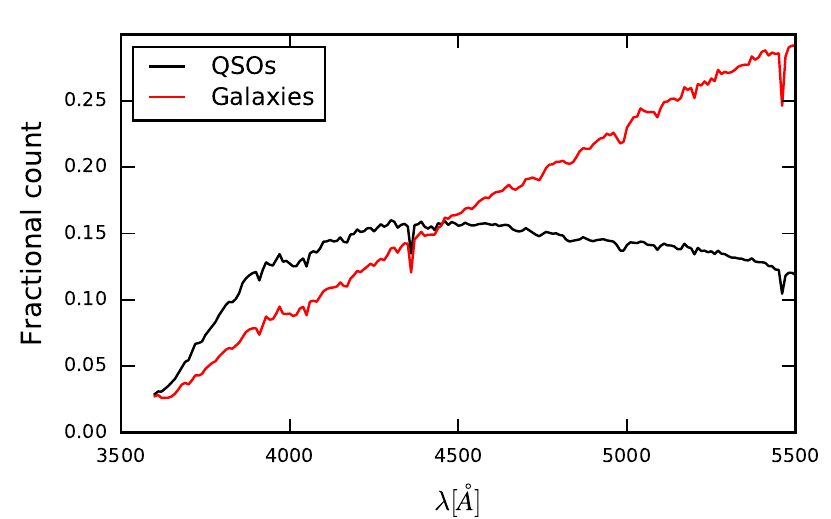}
\caption{Contribution of galaxies vs. QSOs to the ISM spectra. 
The plot shows the fraction of objects with relative absorption in the range \SIrange{0.9}{1.1}{}.
The median is almost always within this range, so the density of objects determines the amount of statistical noise.
At wavelengths shorter than $\unsim\SI{4400}{\angstrom}$ the median value is determined mostly by QSOs.
\label{fig:qso_vs_gal}}
\end{figure}

Since, generally speaking, QSOs are blue and galaxies are red, the blue end of the median spectra is dominated by QSOs (\cref{fig:qso_vs_gal}), and the red by galaxies, where their much larger number contributes in addition to their colour. The split point occurs at roughly \SI{4400}{\angstrom} which corresponds to a redshift of $\unsim2.6$.
The \LymanA{} autocorrelation data spans $2<z<3.5$, but most of the statistical weight is near $z\sim 2.2$. This means that for the most part, our ISM estimation relies on the QSO spectra, which include the \LymanA{} QSOs and lower redshift QSOs, and we cannot leverage significantly the power in the galaxy spectra. \LymanA{} QSOs account for about half of the  flux from QSOs in the $u$ and $g$ bands.

\begin{figure*}
\renewcommand{\AA}{\text{\r{A}}}
\includegraphics[width=\linewidth]{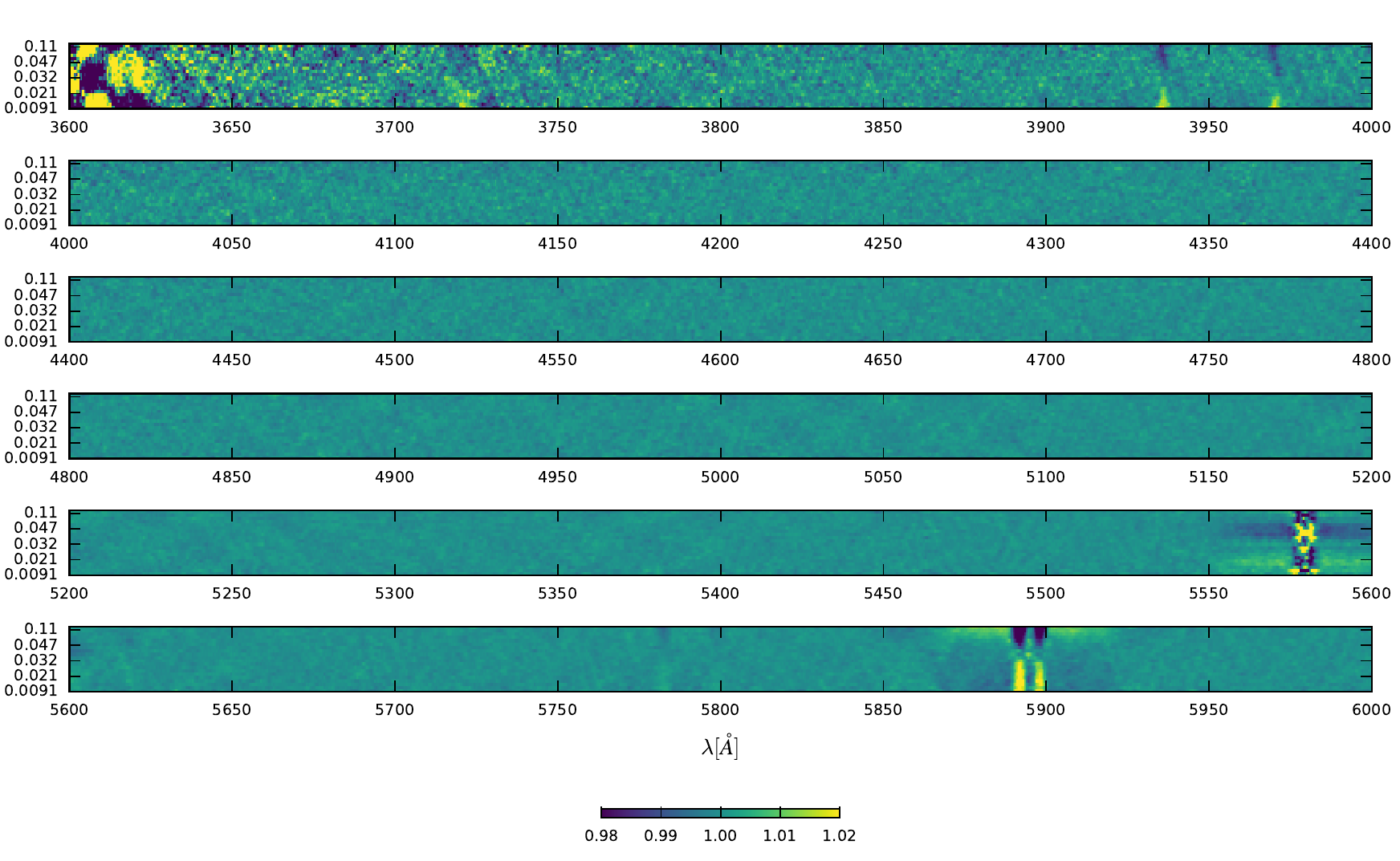}
\caption{Stacked ISM spectra from BOSS galaxies and QSOs, in bins of extinction. The bin boundaries are chosen so that there is a similar number of spectra in each bin.
The mean absorption at each redshift has been removed to show only how absorption correlates with extinction.
Our range of interest is \SIrange{3600}{5500}{\angstrom} (\LymanA{} redshift of \numrange{1.95}{3.5}).
The blue extreme is noisy so we cannot say much about it (it also has small contribution to the \LymanA{} autocorrelation due to the weighting scheme).
The \CaHK{} lines (\SIlist{3968;3933}{\angstrom}) are clearly the most notable feature in our range.
The artifact at $\unsim\SI{5580}{\angstrom}$ is caused by the transition between the red and blue spectrographs.
The \ion{Na}{i} doublet can be seen at \SI{5890}{\angstrom} and \SI{5896}{\angstrom}.
Also visible are the DIBs at \SI{5780}{\angstrom} and, to a lesser extent,  \SI{5797}{\angstrom}. In order to show small changes in intensity, we use a colour scale that saturates some of the data. 
\label{fig:ism_detail}}
\end{figure*}

The stacked ISM spectra are shown in \cref{fig:ism_detail}. Each row represents a spectrum in a single extinction bin. The mean absorption has been removed to show only features that are correlated with extinction. One can see that the spectra are very noisy at short wavelengths, where there is a penury of photons, and that only very strong lines appear. The \CaHK{} lines (\SIlist{3968;3933}{\angstrom}) are clearly the most notable feature in our range. The artifact at $\unsim\SI{5580}{\angstrom}$ is caused by the transition between the red and blue spectrographs.
The \ion{Na}{i} doublet can be seen near \SI{5890}{\angstrom}.
Also visible are the DIBs at \SI{5780}{\angstrom} and, to a lesser extent,  \SI{5797}{\angstrom}.

\section{Estimating the ISM influence on the BAO signal}
\label{sec:ism_estimate}
We proceed to estimate the influence of the ISM absorption lines on the estimated baryonic peak. We identify two potential ways for the ISM to produce a BAO-like feature in the autocorrelation function:
The first is correlation between different lines at different wavelengths. Such correlations would mimic intrinsic \LymanA{} correlations in the parallel direction, along the line of sight, and have no angular dependence. 
Additionally, there could be angular correlations in all (or some) of the line strengths, due to preferred scales in the MW gas distribution. 

\subsection{Wavelength/parallel component}\label{s:wave}

For correlations between lines to bias the BAO measurement, lines would need to have wavelength separations that correspond to a line-of-sight comoving distance of the same order as the baryon peak. The spurious feature created by such lines would be at a comoving distance that is the mean of the apparent comoving distance of the lines (when wrongly interpreted as \LymanA{}).

To create a BAO-like signal of the same strength ($\unsim\SI{5e-5}{}$) using absorption line pairs, we would need relative absorption which is the square root of the signal strength, $\unsim\SI{7e-3}{}$, for a particular redshift value.
However, because we average over a large redshift range, multiple pairs of different line transitions would have to conspire to create a single feature that would not be smoothed.

Consider two absorption lines with effective profile widths $\Delta \lambda_1$ and $\Delta \lambda_2$ and average relative absorptions $f_1, f_2$, over a redshift range $\Delta \lambda_\text{survey}$. For simplicity, we treat the measured absorption profiles as boxcar functions. In addition, for simplicity, we disregard the nonlinear relation between wavelength/redshift and comoving distance.

Assuming the correlation function resolves the lines (in comoving space), their contribution (without taking noise into account) would be:
\begin{equation}
f_1 f_2 \frac{\Delta \lambda_1 \Delta \lambda_2}{(\Delta \lambda_\text{survey})^2}.
\end{equation}

The effect is attenuated by the fact that we measure absorption relative to the mean transmittance over the whole survey. On the other hand, the statistical weight contribution varies with wavelength.
Therefore, it would be wise to take a smaller, more conservative estimate
for $\Delta \lambda_\text{survey}$.
For BOSS, most of the data is around redshift \SIrange{2.1}{2.5}{}, so the wavelength range is about \SI{500}{\angstrom}.

For the strongest lines in our range of interest, the \CaHK{}, we find in the stacked ISM spectra a relative absorption of \SI{10}{} percent, but less than \SI{3}{} percent change in absorption between the highest and lowest extinction bins. Using a rough value of \SI{5}{} percent, we obtain:

\begin{equation}\label{eq:ca}
0.05^2 \frac{(\SI{5}{\angstrom})^2}{(\SI{500}{\angstrom})^2} = \SI{2.5e-6}{}
\end{equation}

This number is 20 times smaller than the BAO amplitude of $\unsim\SI{5e-5}{}$, however it assumes that our stacks based on the dust maps can recover the full potential correlation between the \CaHK{} lines.

We estimate the bias that could be introduced to the BAO peak estimation by ISM lines. As we show in \cref{s:stacks}, the current data, especially in the bluest part of the spectrum, are insufficient to clearly detect most lines, so we cannot rule out hypothetical correlations from lines that we know exist in this wavelength range (e.g., \citealt{morton03}). As a proxy for these lines, we assume the amplitude of the \CaHK{}-induced correlation (\cref{s:dist_bins}), but allow it to shift in apparent comoving distance, in order to simulate the potential effect of ISM lines which may remain hidden in our noise. We fit a Gaussian curve to the sum of the real and spurious signals to find the displacement in the primary peak position. We find that the maximum bias is obtained when the lines are positioned $\unsim\SI{0.7}{\sigma}$ from the primary peak, where $\sigma$ is the standard deviation of the BAO signal. We obtain a value of about \SI{0.3}{} percent to the BAO peak displacement, if indeed the ISM correlation is about 20 times smaller than the cosmological signal. Specifically for the \CaHK{} lines, the maximum displacement, of order \SI{0.05}{} percent, occurs in the $0.1<\mu<0.5$ angular wedge. The baryon peak has a constant $r$, whereas ISM lines have a constant $\rpar$, which means their relative position depends on $\mu$. The exact bias depends on the details, mainly the location of the spurious peak, and to a certain extent the fit method, but it is likely insignificant for BOSS QSOs.

\subsection{Angular component}\label{s:ang}

\begin{figure} 
\renewcommand{\AA}{\text{\r{A}}}
\includegraphics[width=\columnwidth]{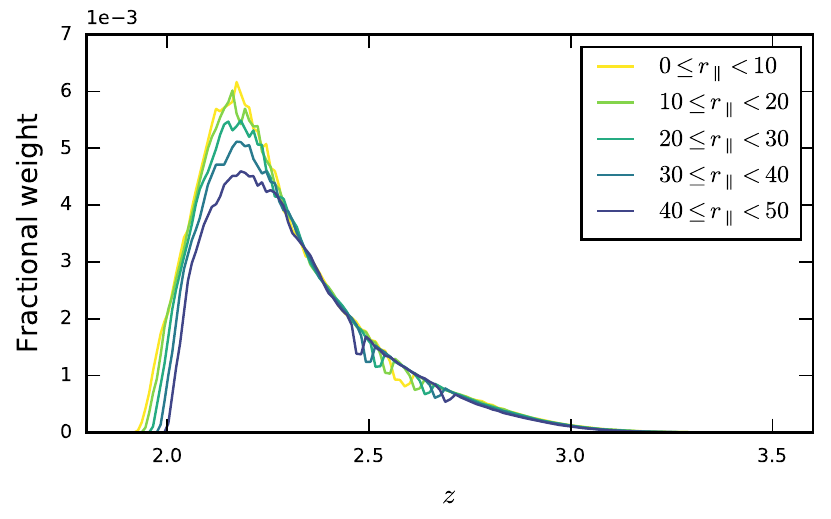}
\caption{Weights of pixel pairs in the autocorrelation function, in slices of redshift, binned by their parallel separation $\rpar$.
The low redshift end of the \LymanA{} forest contributes only to low separation bins, creating a selection effect.
The dips around $z\sim2.6$ are a result of masking the telluric \ion{Hg}{i} line.
\label{fig:weights_parallel}}
\end{figure}

\begin{figure} 
\renewcommand{\AA}{\text{\r{A}}}
\includegraphics[width=\columnwidth]{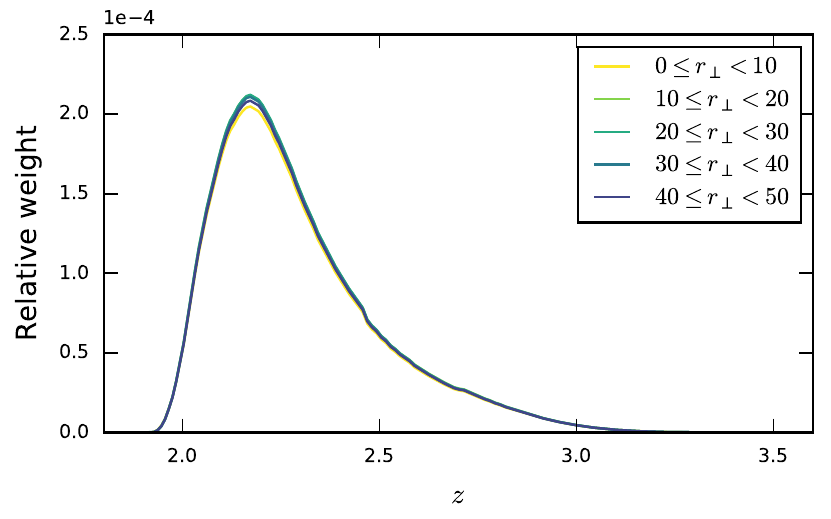}
\caption{Weights of pixel pairs in the autocorrelation function, in slices of redshift, binned by their transverse separation $\rtrans$.
Because the number of pairs scales linearly with the transverse separation, we divided the average scale from each bin.
It is clear to see there is very little dependence on the transverse separation, even at high redshifts where the density of QSOs is the smallest.
\label{fig:weights_transverse}}
\end{figure}

Most of the statistical weight in BOSS data occurs around $z=2.2$. This can be seen in  \cref{fig:weights_parallel,fig:weights_transverse}, where we calculate the relative contribution of pixel-pairs at different redshifts to the autocorrelation function. The baryon peak scale of \SI{100}{\hrmpc} corresponds to about \SI{1.5}{\degree} at this redshift. As a consequence of the rather narrow peak, having pairs at different redshifts might not be enough to completely erase transverse correlations that may exist at the $\sim$\SI{1}{\degree} scale in the ISM. The sky density of galaxies and QSOs in SDSS is insufficient to create ISM stacks at every position, and quantify this. Instead, we can use extinction maps as a proxy to ISM absorption lines, assuming the absorption lines are correlated with dust column (which we know is only true to first order, e.g.,  \citealt{2015MNRAS.447..545B}).

\begin{figure} 
\renewcommand{\AA}{\text{\r{A}}}
\includegraphics[width=\columnwidth]{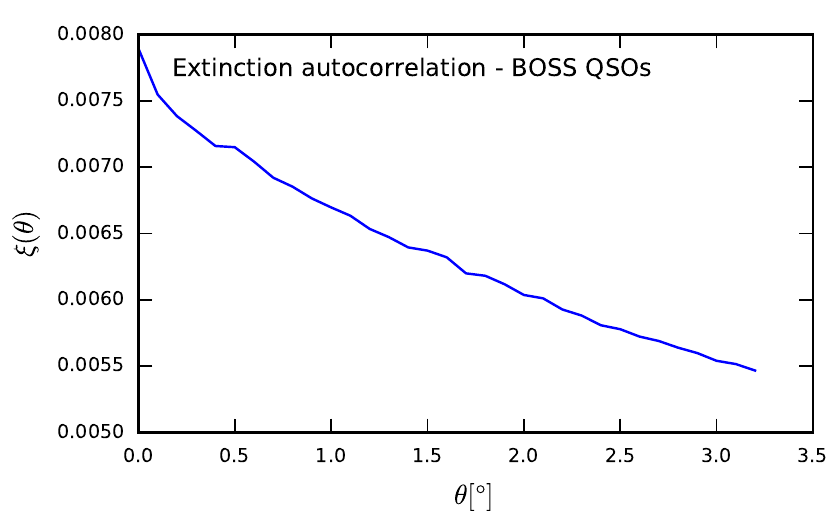}
\caption{Autocorrelation of extinction as a function of the angle of separation $\theta$.
Produced using the extinction values from SDSS for every QSO, summed over all pairs of QSOs.
The SDSS extinction value is derived from the SFD maps \citep{1998ApJ...500..525S}.
\label{fig:corr_extinction_boss}}
\end{figure}

We calculate the angular autocorrelation of extinction in two ways:
In Method 1, we use the foreground extinction through every QSO sightline using the maps of \citet{1998ApJ...500..525S}.
We compute the autocorrelation in angular bins for all possible pairing within a maximum separation of $\unsim\SI{3}{\degree}$. The resulting angular autocorrelation can be seen in \cref{fig:corr_extinction_boss}. Clearly, there is no preferred scale, as previously shown in Figure 9 of \citet{1998ApJ...500..525S}. 

\begin{figure} 
\renewcommand{\AA}{\text{\r{A}}}
\includegraphics[width=\columnwidth]{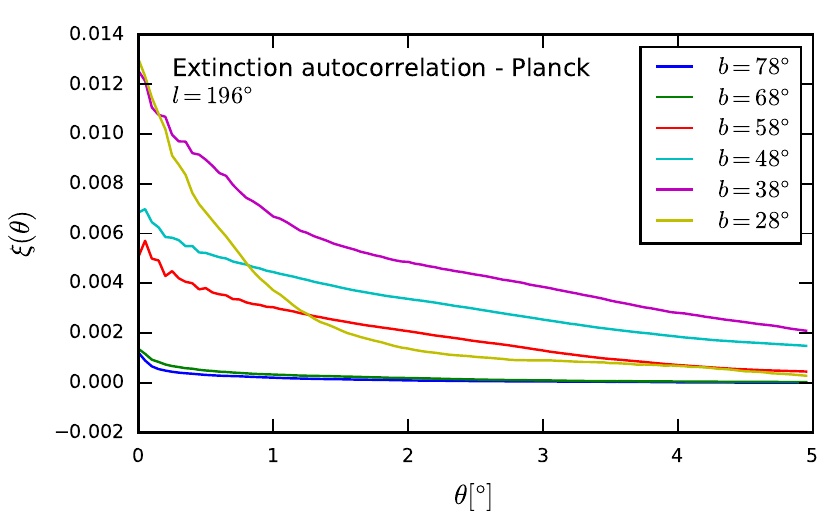}
\caption{Same as \cref{fig:corr_extinction_boss}, 
but produced using extinction maps from \citet{2016A&A...586A.132P}.
The various curves show the autocorrelation inside circular discs ($r=\SI{10}{\degree}$) at different galactic latitudes.
\label{fig:corr_extinction_planck}}
\end{figure}

In Method 2, we use the more recent extinction map from the Planck collaboration \citep{2016A&A...586A.132P}. We calculate the angular autocorrelation for different regions of the sky, within the SDSS field. The result, as seen in \cref{fig:corr_extinction_planck}, behaves similarly to the one obtained with Method 1. While extinction varies greatly in magnitude at different galactic latitudes, the autocorrelation function in all fields has a smooth slope which decreases with angular separation. It appears there is no preferred scale in the dust column density.
We therefore do not expect any significant transverse spurious correlation from the ISM. However, this is only a first order estimation, since ISM lines do not correlate perfectly with dust column density \citep{2015MNRAS.447..545B}.

\subsection{Autocorrelation of stacked ISM spectra}\label{s:stacks}

Since the correlation computation is multi-staged, complex, and very non-linear, we proceed with a data driven approach to quantifying the ISM absorption effect on the autocorrelation estimator. Optimally we would want to derive an ISM spectrum along every sightline, however, since most of our signal in the blue is from QSOs, the source density in BOSS is insufficient. Instead, since to first order all the absorption lines correlate with extinction, we can make a reasonable estimate of the effect of ISM absorption, by using extinction as a proxy for line strength. 

For every QSO sightline we use the dust extinction (from the maps of \citealt{1998ApJ...500..525S}) to substitute the actual spectrum and  \LymanA{} forest with an ISM spectrum stacked from sightlines with similar extinctions. The baryon peak pipeline then proceeds normally, removes the mean transmittance as before, and only features that are correlated with extinction remain.

\begin{figure}
\renewcommand{\AA}{\text{\r{A}}}
\begin{center}
\includegraphics[width=\columnwidth]{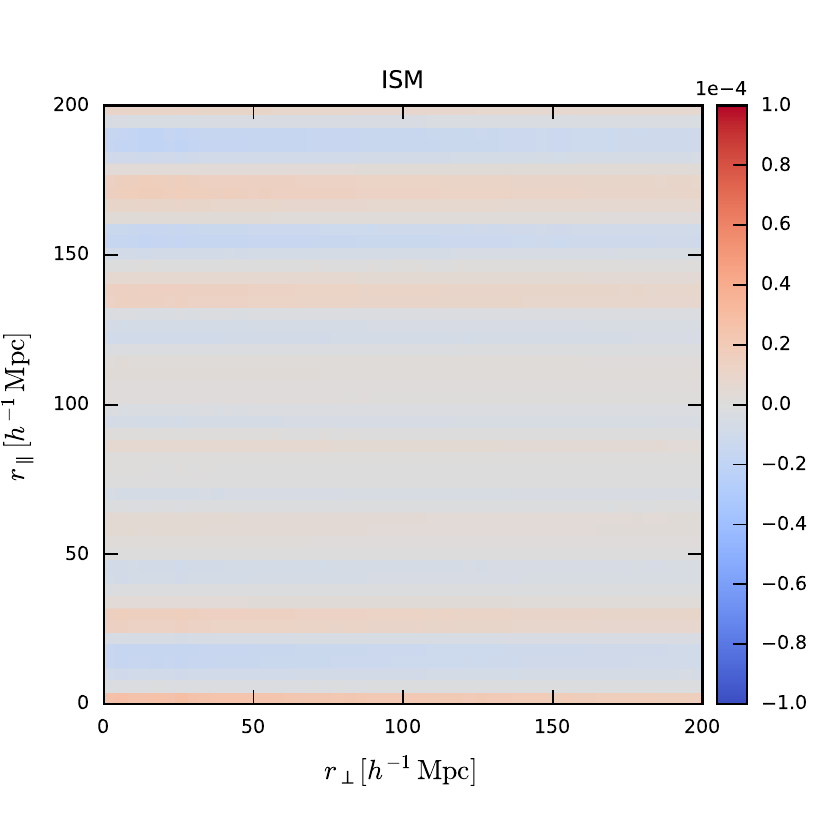}
\end{center}
\caption{ISM autocorrelation $\hat{\xi}_A$ of the change in transmittance $\delta_F$, in comoving distance bins in parallel and transverse directions. 
Note that the colour range is four times smaller than in \cref{fig:corr_2d}. Clearly any correlation the ISM might introduce is much smaller than the cosmological signal. Since we use stacked spectra based on dust maps we are not sensitive to correlations that might exist due to ISM cloud sizes that could potentially break the stripes in the $\rtrans$  direction.\label{fig:corr_2d_ism}}
\end{figure}

\begin{figure}
\renewcommand{\AA}{\text{\r{A}}}
\includegraphics[width=\columnwidth]{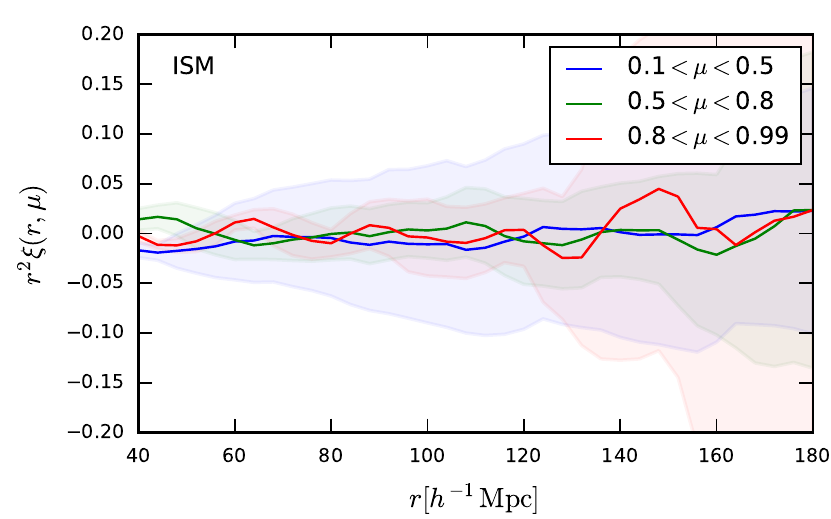}
\caption{ISM autocorrelation multiplied by distance squared in three angle ranges where $\mu$ is the cosine of the angle of the coordinate $(\rpar,\rtrans)$.
As in \cref{fig:corr_2d_ism} there is effectively no signal in the transverse wedge (small $\mu$). Note the different scales on the Y-axis compared with \cref{fig:corr_mu}.
\label{fig:corr_mu_ism}}
\end{figure}

We then compute the autocorrelation function in bins of comoving separation. We discard pixels pairs with $z<2.1$ because the ISM spectra are dominated by noise in this range. We plot the ISM autocorrelation estimator in a similar way to the \LymanA{} forest, in \cref{fig:corr_2d_ism,fig:corr_mu_ism}.

From this calculation we find the following. First, the amplitude of the signal from the ISM is about an order of magnitude weaker than the cosmological component. Secondly, the amplitude of the correlation drops by a factor of about 1.5 with $\rtrans$, with no preferred scale, as obtained before from dust maps (\cref{fig:corr_extinction_boss} and \cref{fig:corr_extinction_planck}). Because we use stacks based on extinction, we cannot see changes along $\rtrans$ that would occur due to potential clustering of lines, as measured by \citet{2015MNRAS.447..545B}, hence the straight lines. Instead we are mostly sensitive to correlations between different absorption lines that appear in $\rpar$. The \CaHK{} doublet mimics a \SI{27}{\hrmpc} separation, is probably responsible for the positive correlation around this value of $\rpar$, and we confirm this in the following section. These results are consistent with our  indirect analysis in \cref{s:wave} and \cref{s:ang}.

\subsection{Autocorrelation in distance bins}\label{s:dist_bins}
In order to study the effect of specific lines, most notably \CaHK{}, and trace the source of the stripes seen in \cref{fig:corr_2d_ism}, we divided the data into bins of parallel comoving distance (effectively redshift or wavelength bins). 

\begin{figure*} 
\renewcommand{\AA}{\text{\r{A}}}
\includegraphics[width=\linewidth]{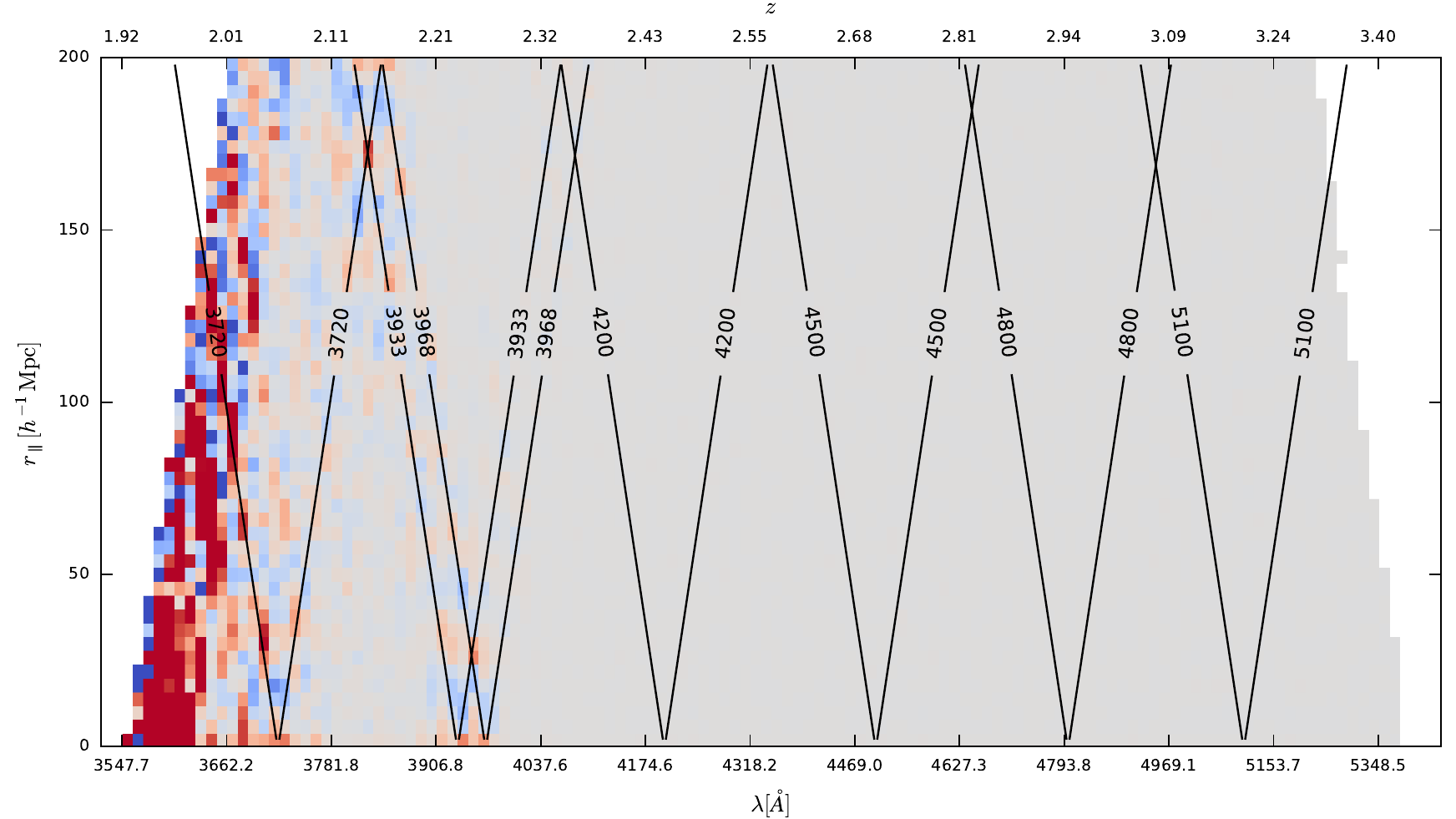}
\caption{Autocorrelation of ISM spectra as a function of (apparent) parallel comoving separation.
We use BOSS QSO sightlines where every QSO's \LymanA{} forest was replaced with estimated ISM absorption based on the QSO's extinction.
Each pixel comprises contribution from two different wavelengths.
Each V-shaped curve shows the contribution of a single wavelength.
Note the peak at the intersection of the two \CaHK{} lines (\SIlist{3968;3933}{\angstrom}). 
The blue extreme, including an apparent feature at $\unsim\SI{3720}{\angstrom}$ that could be due to atomic iron, is noisy so we cannot say much about it (it also has small contribution to the \LymanA{} autocorrelation due to the weighting scheme).
The colour scale is \SI{\pm 4e-4}{} which is comparable to the scale of the baryon peak, but features that affect a single distance slice are reduced by averaging over all distance slices.
\label{fig:ism_dist_bins}}
\end{figure*} 

\Cref{fig:ism_dist_bins} shows how the \CaHK{} lines create spurious correlations with each other as well as with other weaker features in the ISM spectrum, including a feature at $\unsim\SI{3720}{\angstrom}$, which we barely detect, nor can we constrain its carrier (it could perhaps be atomic iron, see \citealt{morton03}). These \CaHK{} correlations contribute to the strongest stripes in \cref{fig:corr_2d_ism}, near \SI{30}{\hrmpc}, \SI{140}{\hrmpc}, and \SI{170}{\hrmpc}. The total correlation contribution of the \CaHK{} peaks, weighted according to their corresponding redshift bin, is about \SI{6e-6}{}, comparable to our estimate in \cref{s:ang}, and an order of magnitude smaller than the BAO signal. Since the largest impact is from the \CaHK{} lines, our previous use of the contribution from these lines to the autocorrelation as proxies for spurious signal seems justified.

\section{Impact on future Surveys}

While we find that the ISM currently has a negligible impact on the cosmological signal recovered from BOSS QSOs, we would like to quantify whether future surveys, such as the Dark Energy Spectroscopic Instrument (DESI), need to take this contaminant into consideration. DESI will obtain many more spectra, with a resolving power similar to that of BOSS, as well as a comparable SNR, with the larger aperture telescope compensating for the fainter QSOs \citep{2016arXiv161100036D}. 

We use the QSO density estimate from the DESI final design report \citep[Table 2.7]{desi-fdr-science}  to simulate a random \LymanA{} forest which resembles what we expect from DESI, but without any large scale correlations.
We choose \LymanA{} forests from random SDSS QSOs with a survey area of about \SI{10000}{\square\deg}.
This value lies between the proposed descoped DESI survey (\SI{9000}{\square\deg}) and the full DESI survey (\SI{14000}{\square\deg}).

We begin by taking the expected DESI QSO density as a function of redshift and using it as a probability distribution function (PDF).
We smooth the PDF using linear interpolation.
We draw \SI{672000}{} redshift values using the smoothed PDF, and match every value with an SDSS QSO at a similar redshift.
We take the pre-calculated $\delta_F(z)$ values of every SDSS QSO, and rescale them to the chosen redshift value.
Since most SDSS \LymanA{} forests will appear more than once in our sample, we multiply every selected QSO by a random sign, to avoid contaminating the autocorrelation function with positive contribution from same-object pairs.
We then draw another random SDSS QSO, and use its sky position, with a small random displacement as the position of our new artificial DESI \LymanA{} forest.
We then compute the autocorrelation function of the new collection of \LymanA{} forests.

While the large scale structure is removed by the randomization process, the resulting autocorrelation bins contain statistical noise which is similar to what we expect from the real DESI survey, under the assumption that the distributions of SNR and instrumental uncertainty values in the \LymanA{} forest are not too different. 
We also obtain the accumulated statistical weight as a function of separation and redshift.

The total number of pixels summed and the total statistical weight are both about 10 times larger than in BOSS.
The standard deviation of individual autocorrelation bins is $\unsim\SI{5e-6}{}$, which is the same order of magnitude as the correlation arising from the \CaHK{} lines, which are representative of the maximal spurious signal. In the near-parallel angular wedge, where a few bins are summed together, the \CaHK{} lines might have a noticeable contribution, but this happens at a separation a few times smaller than the baryon acoustic peak. Overall using our method we find no indication the ISM will significantly influence DESI results. However, we caution that this is under the first-order assumption we had to make that the ISM follows the dust column density and has no additional covariance. 

While \CaHK{} lines could simply be masked by future surveys, this does not resolve correlations arising from other lines or possible structures in the ISM which we cannot probe due to our limited signal in the bluest part of the spectra. A better method to treat these biases would be to empirically derive the ISM spectrum around every QSO sightline, by stacking nearby lower-$z$ spectra in the observer frame, and removing the ISM contribution from every spectrum. As a consequence, \LymanA{} cosmology would indirectly  benefit from a large density of other unrelated sources. 

\section{Conclusions}
By building an independent (though similar) pipeline we reproduce the detection of the BAO signal in the \LymanA{} forest of BOSS QSOs. We analyze in different ways the possible biases that this measurement may suffer from, due to interloping lines from the MW ISM. The spurious signal could be produced by correlations between lines, i.e., in the parallel direction, or by angular correlations on the sky, affecting the transverse direction. 
 
We find that the imprint of ISM absorption is too small to significantly affect current BAO results from the 3-D \LymanA{} forest.  As future surveys such as DESI contain more data with more precision in BAO, it might be preferable to explicitly correct for these effects.

While our current analysis suffers from a lack of signal in the blue wavelength range of interest, these future surveys should have a better handle on the ISM, allowing them, via the methods presented here, to better measure and correct for biases due to the MW ISM.

\section*{Acknowledgements}
We would like to thank the 2nd referee A. Font-Ribera, and K.G. Lee for useful comments on this manuscript, and Roman Garnett for providing us early access to the DLA catalogue from \citet{2016arXiv160504460G}. We further thank Julien Guy for his advice and for making numerical autocorrelation results available to us for comparison. This research was supported by Grant No. 2014413 from the United States-Israel Binational Science Foundation (BSF).
\AckSDSSIII




\bibliographystyle{mnras}
\bibliography{ism_article} 

\bsp	
\label{lastpage}
\end{document}